\documentclass[12pt]{iopart}
\usepackage[pdftex]{graphicx}

\begin{document}

\title[Efficient basis for the Dicke Model II]{Efficient basis for the Dicke Model II: wave function convergence and excited states}

\author{Jorge G. Hirsch and Miguel A. Bastarrachea-Magnani}

\address{Instituto de Ciencias Nucleares, Universidad Nacional Aut\'onoma de M\'exico, Apdo. Postal 70-543, Mexico D. F., C.P. 04510}
\ead{miguel.bastarrachea@nucleares.unam.mx}
\begin{abstract}
An extended bosonic coherent basis has been shown by Chen {\it et al} \cite{Chen08} to provide numerically exact solutions of the finite-size Dicke model. The advantages in employing this basis, as compared with the photon number (Fock) basis, are exhibited to be valid for a large region of the Hamiltonian parameter space and many excited states by analyzing the convergence in the wave functions.
\end{abstract}

\section{Introduction}
The Dicke Hamlitonian describes a system of $\mathcal{N}$ two-level atoms interacting with a single monochromatic electromagnetic radiation mode within a cavity. It is described in the accompanying article \cite{Bas13}. The purpose of this second part is to show that the benefits to employ the coherent basis are valid for a large region of the Hamiltonian parameter space, not only to obtain converged values of the energy, but also for the wave function, for the ground state and for a significative part of the energy spectra. It can be particularly useful to study the presence of chaos \cite{Emary03,Lam05} and of excited states phase transitions \cite{Per101,Per201} in this model.
 

The interaction between a system of $\mathcal{N}$ two-level atoms and a single mode of a radiation field can be described by the Dicke Hamiltonian:
\begin{equation}
H_{D}=\omega a^{\dagger}a + \omega_{0} J'_{z} + \frac{\gamma}{\sqrt{\mathcal{N}}}\left(a+a^{\dagger}\right)\left(J'_{+}+J'_{-}\right).
\end{equation}
The frequency of the radiation mode is $\omega$, which has an associated number operator $a^{\dagger}a$. For the atomic part $\omega_{0}$ is the excitation energy, meanwhile $J'_{z}$, $J'_{+}$, $J'_{-}$, are collective atomic pseudo-spin operators which obey the SU(2) algebra. It holds that if $j(j+1)$ is the eigenvalue of $\mathbf{J}^{2}=J_{x}^{'2}+J_{y}^{'2}+J_{z}^{'2}$, then $j=\mathcal{N}/2$ defines the symmetric atomic subspace which includes the ground state. The interaction parameter $\gamma$ depends principally on the atomic dipolar moment. 


\section{Numerical Diagonalization}
We compare the minimal truncation needed to obtain convergence of the solution, using the two basis defined in Ref. \cite{Bas13}:  the coherent basis $|N;j,m\rangle$ and the Fock basis $|n;j,m\rangle$. 

The wave functions, expanded in the truncated Fock (F) and coherent (C) basis are, for a given $j= {\cal N}/2$:
\begin{equation}
|\Psi^{k}_X\rangle=\sum\limits_{x=0}^{x_{max}} \sum\limits_{m=-j}^{j} C^{k, X}_{m,x} |x;j,m\rangle,
\label{psiF}
\end{equation}
where $x=n$ for $X=F$, and  $x=N$ for $X=C$, and $k=1, ..., (x_{max}+1)(2j +1)$ enumerates the eigenstates ordered by their energies $E^k_X$ with $k=1$ assigned to the ground state.

\subsection{The Wave Functions}

\begin{figure}[h!] 
\centering
\qquad
\includegraphics[width=0.45 \textwidth]{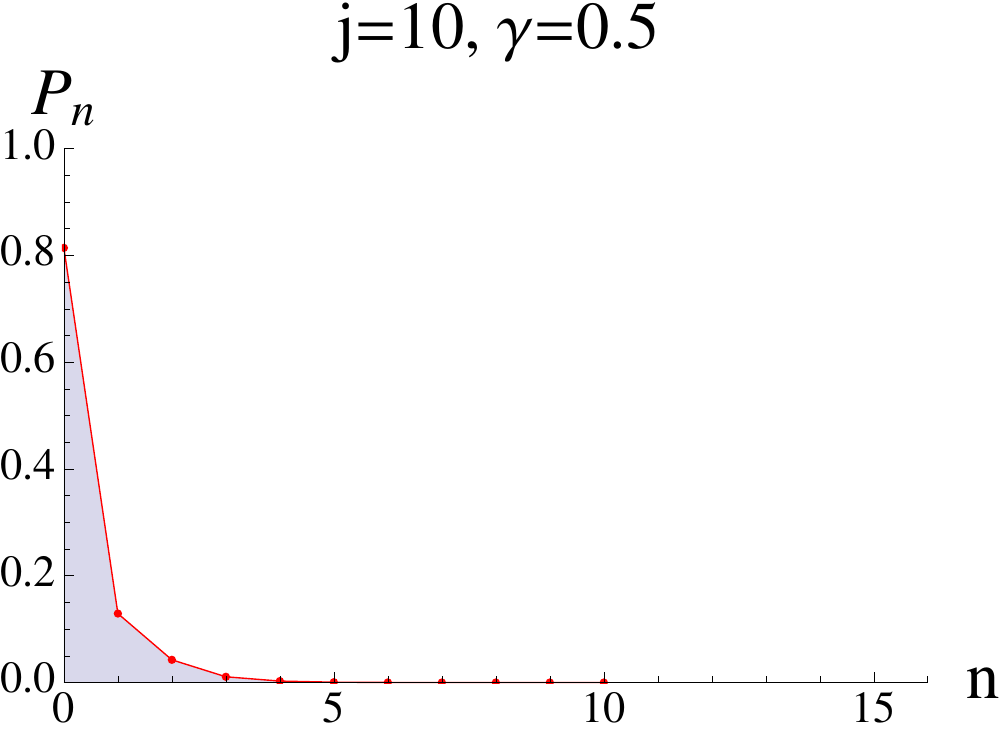}
\includegraphics[width=0.45 \textwidth]{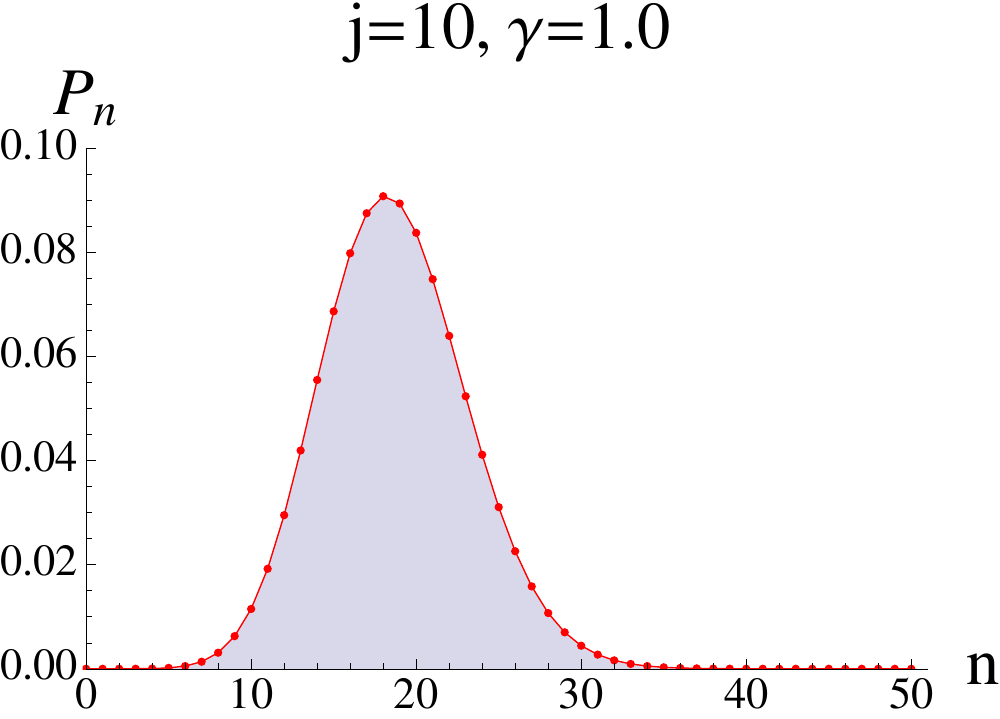}
\caption{$P_{1,n}$ as function of $n$ in the Fock basis, for $j=10, \, \gamma=0.5, \, n_{max}=15$ (left); and $j=10,  \, \gamma=1.0, \, n_{max}=50$ (right).}
\label{fig1}
\end{figure}

\begin{figure}[h!] 
\centering
\includegraphics[width=0.45 \textwidth]{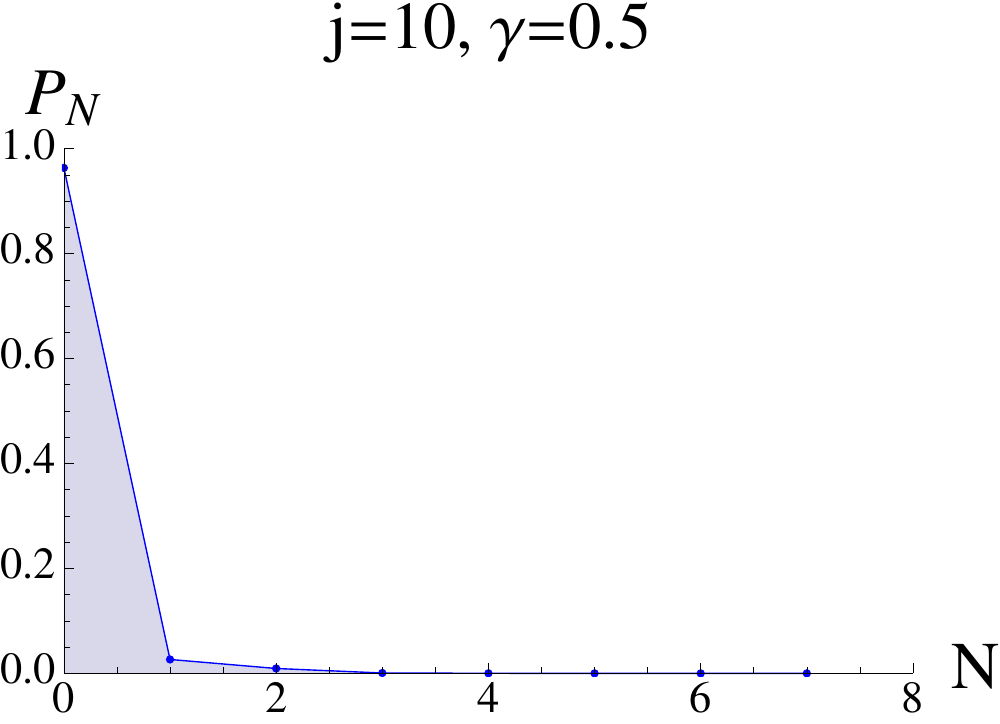}
\includegraphics[width=0.45 \textwidth]{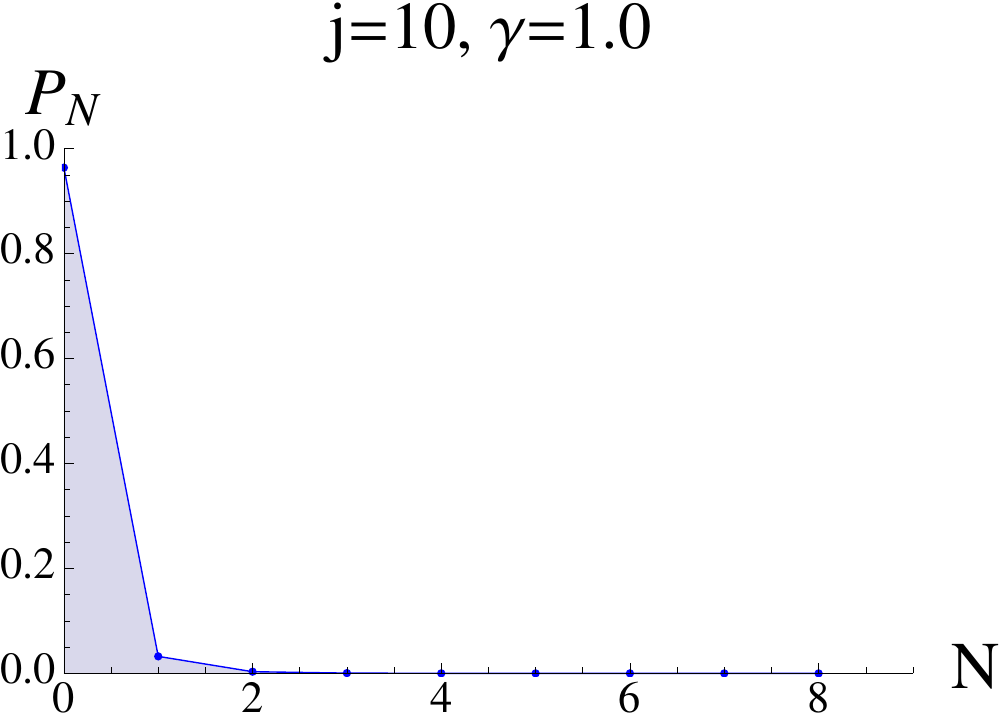}
\caption{$P_{1,N}$ as function of $N$ in the coherent basis, for $j=10, \, \gamma=0.5, \,N_{max}=7$ (left) and $j=10, \, \gamma=1.0, \, N_{max}=8$ (right). }
\label{fig2}
\end{figure}

The probability $P_{n}$ of having $n$ photons in the $k$-th state in the Fock basis, or $P_{N}$ of having $N$ excitations in the coherent basis is:
\begin{equation}
P_{k,x}=|\langle x|\Psi^{k}_X\rangle|^{2}=\sum_{m}|C^{1, X}_{m,x}|^{2}, 
\end{equation}
where $x=n,N$ for $X=F,C$, respectively. The ground state probability distribution $P_{x}=P_{1,x}$ is shown as a function of $n$ or $N$ up to  $n_{max}$ or $N_{max}$, for $\gamma=0.5$ and $1.0$, and $j=10$ in Figs. \ref{fig1} and \ref{fig2}. Both wave functions were calculated with the truncation necessary to have the energy converged with $\Delta E <\epsilon=1\times 10^{-6}$.

From Fig. \ref{fig1} and \ref{fig2} it is clear that many components which contributes very little to the wave function must be included in the calculations to obtain the desired precision in the ground state energy. It can also be observed in the figures that for $\gamma=0.5$, which is $\gamma_c$ in this case, the largest probability is to have no photons in the Fock basis, or no excitations in the coherent basis. The situation is different in the superradiant region, $\gamma =1$, where in the Fock basis the distribution of photons resembles a Gaussian curve, with its maximum at a photon number proportional to the number of atoms, while in the coherent basis the probability of having zero excitations remains dominant. 
This is the power of the coherent basis, which allows to obtain numerically exact ground state wave functions for numbers of atoms which are intractable in the Fock basis.

To study the convergence in the wave function we define its precision  $\Delta P_{X}$ \cite{Basta11}
as
\begin{equation}
\Delta P_{X} \leq  \sum\limits_{m=-j}^j \left|C^{1,X}_{x_{max}+1,m}(x_{max}+1)\right|^2. 
\end{equation}
where $x=n,N$ for $X=F,C$, respectively. This $\Delta P$ criteria demands less computing resources than the $\Delta E$ criteria \cite{Basta11,Bas13}, because it requires only the information about one truncation value ($x_{max}$) instead of two.

\begin{figure}[h!] 
\centering
\includegraphics[width=0.45 \textwidth]{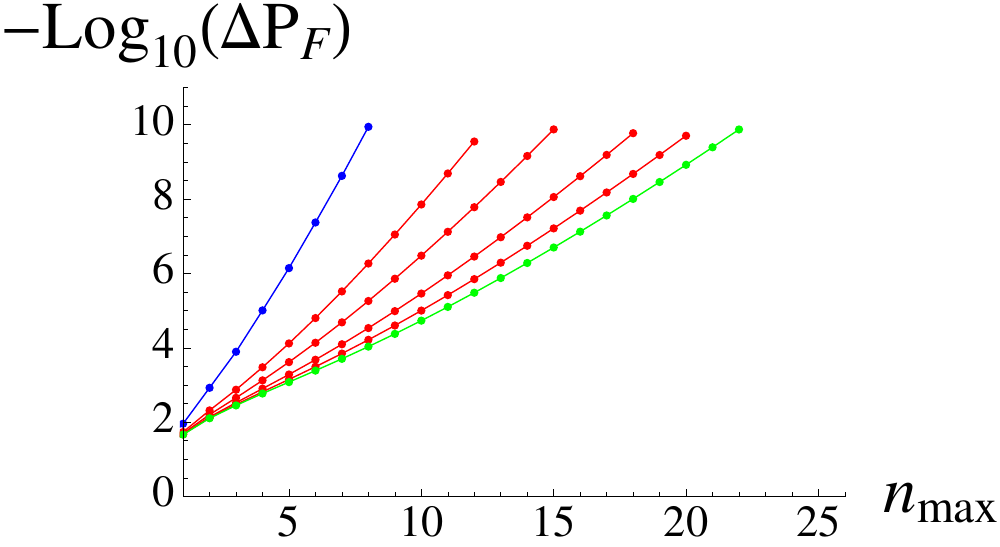}
\includegraphics[width=0.45 \textwidth]{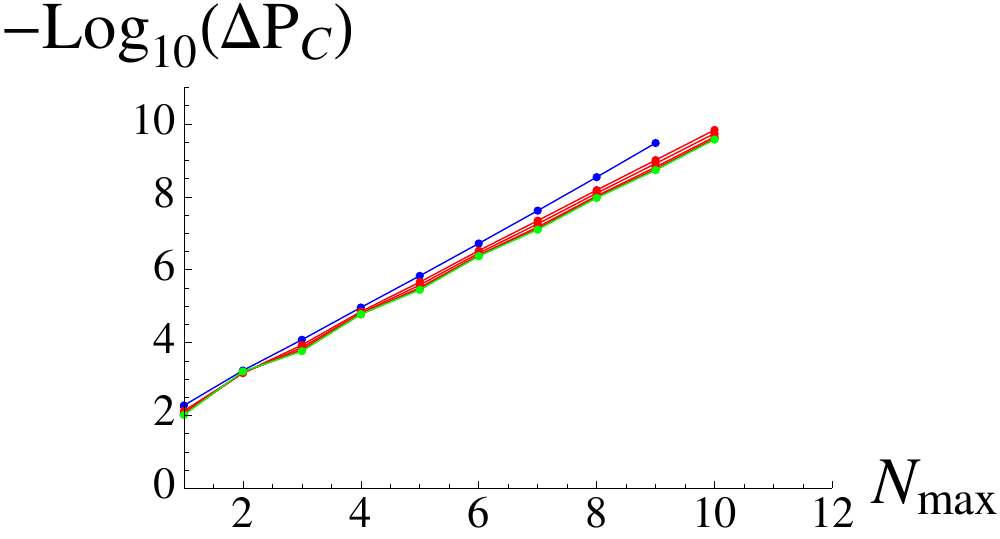}
\caption{(Color online). $\Delta P$ as function of $n_{max}$ (left) and $N_{max}$ (right). From left to right $j=1$ (blue), $5$, $10$, $20$, $30$ and $40$ (green). For $\gamma=0.5$ in resonance. }
\label{fig3}
\end{figure}

Fig. \ref{fig3} displays the plots of $-Log_{10}( \Delta P_{F} )$ as a function of $n_{max}$, and of $-Log_{10}( \Delta P_{C} )$ as a function of $N_{max}$. 

A linear fit, for $j=40$ give us the following relation between $N_{max}$ and $\Delta P_{C}$:
\begin{equation}
-Log_{10}\left(\Delta P_{C}\right)=1.45+ 0.811 \,N_{max} \,\,
\Rightarrow \,\,\, \Delta P_{C}=0.0354\,\,10^{-0.811 \,N_{max}}
\end{equation}


 \section{Numerically exact results for Excited States}
 
In this section we extend the analysis to the excited states. To accurately evaluate a significative part of the energy spectrum is a necessary ingredient in the study of quantum chaos \cite{Emary03} and of excited state quantum phase transitions (ESQPT) \cite{Per101,Per201}.

\begin{figure}[tbp]
\centering
\includegraphics[scale=0.7]{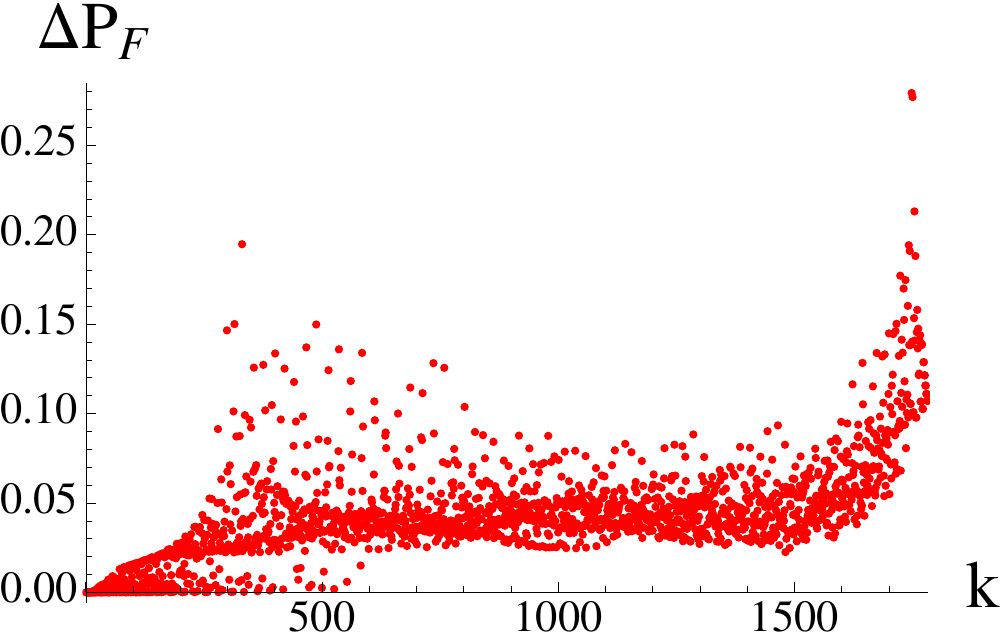}
\includegraphics[scale=0.7]{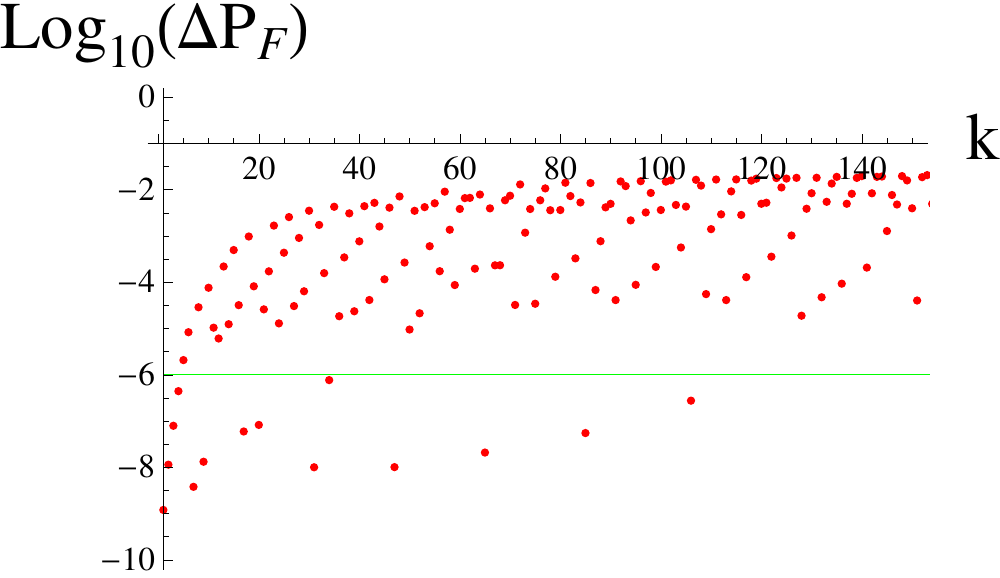}
\qquad
\includegraphics[scale=0.7]{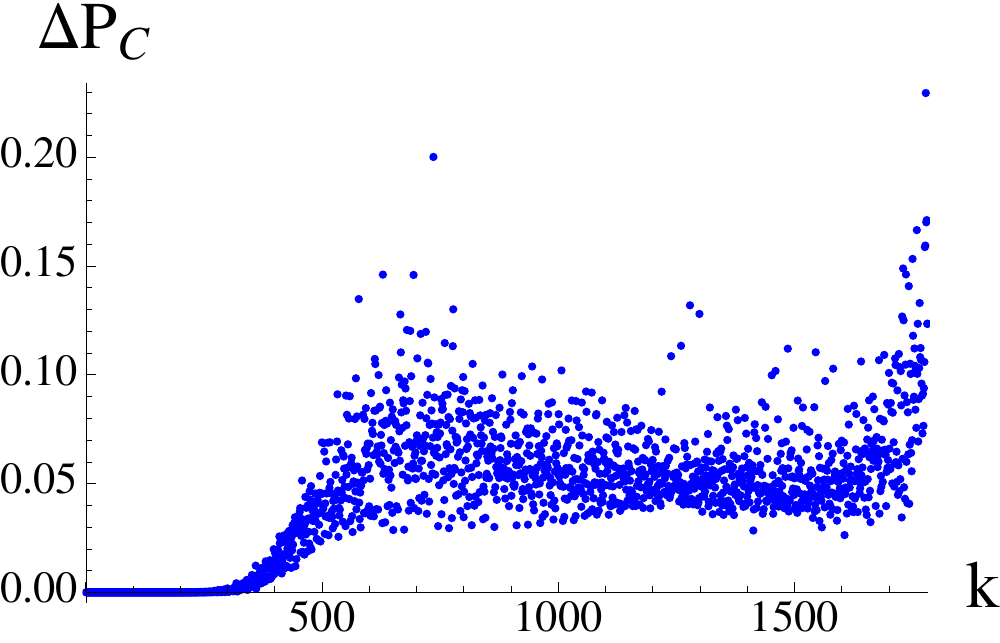}
\includegraphics[scale=0.7]{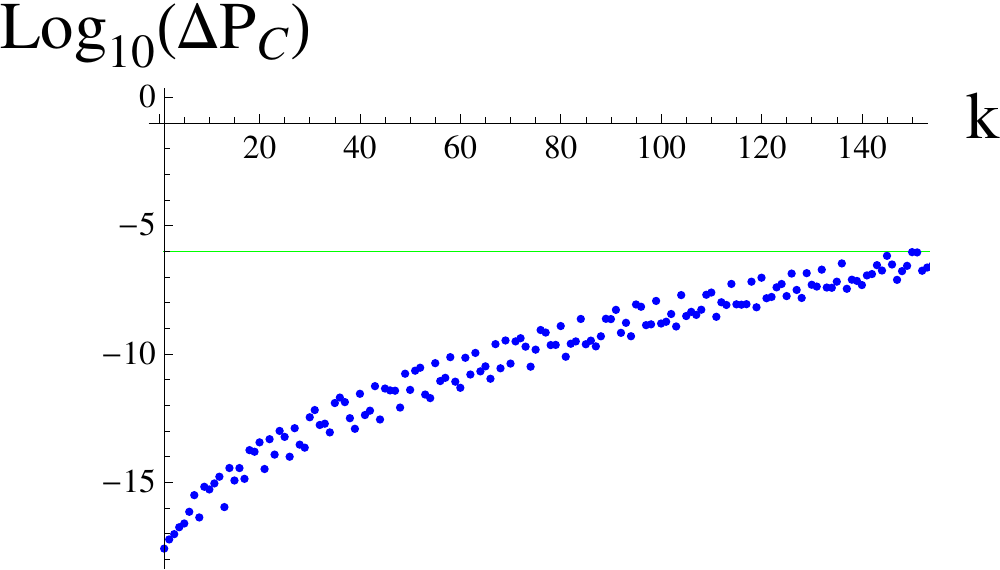}
\caption{$\Delta P$ of all states as a function of the state number. Details are given in the text.}
\label{fig4}
\end{figure}

In figure \ref{fig4} we display plots of $\Delta P$ as a function of the state $k$, for $j=40$, $\gamma=0.5$, $\omega_{0}=1.0$, $N_{max}=20$ and $\epsilon=1\mbox{x}10^{-6}$. 
In the upper figures we show the $\Delta P_{F}$  and in the lower ones  $\Delta P_{C}$.
On the left the vertical scale is linear and all states are listed in the horizontal axis, while on the right hand side the vertical scale is logarithmic and only the 150 states with lower energies are included. The horizontal green line depicts the tolerance $\epsilon$.

\begin{figure}[tbp]
\centering
\includegraphics[scale=0.7]{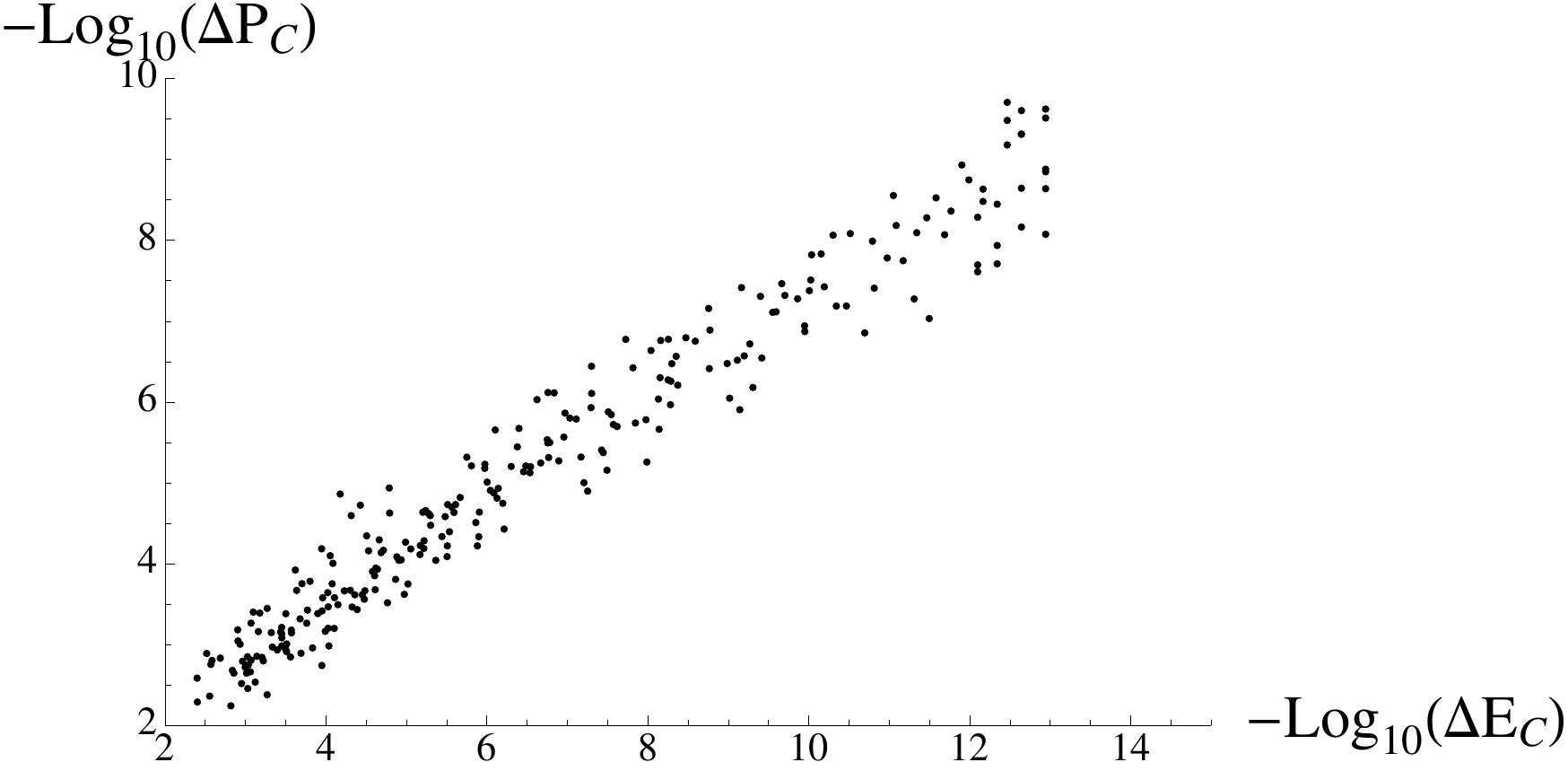}
\caption{$\Delta P_{C}$ vs $\Delta E_{C}$ for the first 250 states, which convergence is under $\epsilon_{2}=1\times 10^{-4}$.}
\label{fig5}
\end{figure}

It is indeed remarkable to observe in Fig.  \ref{fig4} that a few hundred states calculated in the coherent basis have their wave function converged, and $\Delta P_C$, for these states, grows in a smooth and nearly monotonous way as a function of the $k$ index. This is not the case in the Fock basis, where $\Delta P_F$ fluctuates by orders of magnitude between a given state and the following one. It is worth to compare the convergence criteria based in the wave function and described above, with the more standard convergence in energy, which was described in the previous article \cite{Bas13}. 
In figure \ref{fig5} we show $\Delta P_{C}$ versus $\Delta E_{C}$ for the first 250 excited states, $k$, whose energies converged in the coherent basis with $\Delta E < 1\times 10^{-4}$.
 
A linear fit of these data results in
\begin{eqnarray}
&-Log_{10}\left[\Delta P_{C}(k)\right]=0.71077 -  1.10337\,\,Log_{10}\left[\Delta E_{C}(k)\right] \\
&\Rightarrow \,\,\, \Delta P_{C}=0.19464\,\, \left[\Delta E_{C}(k)\right]^{   1.10337}  \nonumber
\end{eqnarray}
The number of states whose $\Delta P$ is smaller than a tolerance $\epsilon$ for $j$ and $n_{max}$ given for the Fock basis and $N_{max}$ for the coherent basis is presented in Table \ref {tab2:}. The two tolerances selected are $\epsilon_{1}=1\times 10^{-6}$ and $\epsilon_{2}=1\times 10^{-4}$, with $\gamma=0.5$ and $\omega_{0}=1$.

\begin{table}[h!]
\centering
\scalebox{0.9}{
\begin{tabular}{|c|c|c|c|c|c|c|c|} \hline
\multicolumn{2}{|c|}{} & \multicolumn{2}{|c|}{$\epsilon_{1}$} & \multicolumn{2}{|c|}{$\epsilon_{2}$} \\ \hline
$j$ & $n_{max}/N_{max}$ & Fock & coherent & Fock & coherent  \\ \hline
10 & 10 & 1 & 18 & 4 & 37 \\ \hline
10 & 15 & 7 & 55 & 15 & 91 \\ \hline
10 & 20 & 20 & 112 & 39 & 166 \\ \hline \hline 
20 & 10 & 0 & 21 & 2 & 43 \\ \hline
20 & 15 & 3 & 65 & 8 & 106 \\ \hline
20 & 20 & 8 & 136 & 20 & 193 \\ \hline \hline
40 & 10 & 0 & 23 & 0 & 48  \\ \hline
40 & 15 & 1 & 70 & 4 & 131  \\ \hline
40 & 20 & 4 & 154 & 12 & 241 \\ \hline
\end{tabular}}
\caption{Number of states whose $\Delta P$ is less than a tolerance $\epsilon$ for $j$ and $n_{max}$ given for the Fock basis and $N_{max}$ for the coherent basis. Tolerances $\epsilon_{1}=1\mbox{x}10^{-6}$ and $\epsilon_{2}=1\mbox{x}10^{-4}$, $\gamma=0\mbox{.}5$, $\omega_{0}=1$.}
\label{tab2:}
\end{table}

The advantages associated with the use of the coherent basis are even more clear in this case, because the number of states whose wave function has converged with the selected tolerance is larger than those whose energies have converged. It should be mentioned, however, that the tolerances in $\Delta P$ are absolute, because its best case value of a fully converged state is zero, and the worst situation, for completely different wave functions,  is one. On the contrary, the energy scale is arbitrary, and can have positive and negative values, even some levels with energies very close to zero. It makes the use of the relative error employed in Ref \cite{Chen08,Basta11} dangerous when the reference energy is very small. But for excited states a fixed value of $\epsilon$ implies the need of more precise digits in the calculated energy, making more difficult the convergence for higher energies. For this reason our $\Delta E$ criteria is more stringent than the $\Delta P$ one: every excited state with converged energy has guaranteed the convergence of its wave function. As the coherent basis provides many converged states with a single truncation value, it is promising to study the presence of Excited States Quantum Phase Transitions (ESQPT), predicted in Dicke-like systems and spin systems for $\gamma$ values deeply in the superradiant phase \cite{Per101,Per201}.

\section{Conclusions}

To obtain the eigenvalues and eigenvectors of the Dicke Hamiltonian for a finite number of atoms it is necessary to perform a numerical diagonalization, employing a truncated boson number space. Two basis, associated with the two integrable limits of the Hamiltonian, are used 
along this work. 
In the present article we have shown that, in most of the Hamiltonian's parameter regions including the QPT, the coherent basis requires a significative smaller truncation. We extended the analysis to the convergence in the wave function, exhibiting both convergence criteria as equivalent, and presented the numerical relationships between them. The study of the probability distributions of the number of bosons was helpful in understanding the differences between the two basis, and the advantages of the coherent basis. The convergence of the energies and the wave functions was also investigated for the excited states, showing that the coherent basis is very powerful also in this case, allowing to obtain hundreds of converged states with a single truncation value. This findings can be very useful in order to observe the presence of quantum chaos around the phase transition, as well as to study the excited states quantum phase transitions.

We thank O. Casta\~nos, R. L\'opez-Pe\~na and E. Nahmad for many useful and interesting conversations.This work was partially supported by CONACyT-M\'exico  and PAPIIT-UNAM 102811.


\end{document}